# Quasi-uniaxial pressure induced superconductivity in stoichiometric compound UTe$_2$

Chongli Yang[1]*, Jing Guo[1,7]*, Shu Cai[1]*, Yazhou Zhou[1]*, Vladimir A. Sidorov[3], Cheng Huang[1,4], Sijin Long[1,4], Youguo Shi[1,4], Qiuyun Chen[5], Shiyong Tan[5], Qi Wu[1], Piers Coleman[2], Tao Xiang[1,4,6], Liling Sun[1,4,7]†

[1]*Institute of Physics, National Laboratory for Condensed Matter Physics, Chinese Academy of Sciences, Beijing, 100190, China*

[2]*Center for Materials Theory, Rutgers University, Piscataway, New Jersey 08854, USA*

[3]*Vereshchagin Institute of High Pressure Physics, Russian Academy of Sciences, 108840 Moscow, Troitsk, Russian Federation*

[4]*University of Chinese Academy of Sciences, Department of Physics, Beijing 100190, China*

[5]*Science and Technology on Surface Physics and Chemistry Laboratory, Mianyang 621908, China*

[6]*Beijing Academy of Quantum Information Sciences, Beijing 100193, China*

[7]*Songshan Lake Materials Laboratory, Dongguan, Guangdong 523808, China*

The recent discovery of superconductivity in heavy Fermion compound UTe$_2$, a candidate topological and triplet-paired superconductor, has aroused widespread interest. However, to date, there is no consensus on whether the stoichiometric sample of UTe$_2$ is superconducting or not due to lack of reliable evidence to distinguish the difference between the nominal and real compositions of samples. Here, we are the first to clarify that the stoichiometric UT$_2$ is non-superconducting at ambient pressure and under hydrostatic pressure up to 6 GPa, however we find that it can be compressed into superconductivity by application of quasi-uniaxial pressure. Measurements of resistivity, magnetoresistance and susceptibility reveal that the quasi-uniaxial pressure results in a suppression of the Kondo coherent state seen at ambient pressure, and then leads to a superconductivity initially emerged on the *ab*-plane at 1.5 GPa. At 4.8 GPa, the superconductivity is developed in three crystallographic directions. The superconducting state coexists with an exotic magnetic ordered state that develops just below the onset temperature of the superconducting transition. The discovery of the quasi-uniaxial-pressure-induced superconductivity with exotic magnetic state in the stoichiometric UTe$_2$ not only provide new understandings on this compound, but also highlight the vital role of Te deficiency in developing the superconductivity at ambient pressures.

## I. INTRODUCTION

A key issue in condensed matter physics is to understand the evolution of electronic states caused by an interplay of electron interactions and competing order [1-3]. Unconventional superconductivity, which often develops in the vicinity of antiferromagnetic, ferromagnetic or charge ordered states [4-13], is a topic of particular interest. The recently discovered superconductor $UTe_2$ is one such example, in which superconductivity emerges from a correlated electronic state [14-18] and coexists with magnetism under pressure [19]. This material hosts magnetic fluctuations [20-24] and is believed to be one of the three-dimensional bulk topological superconductors [15,25-29] which exhibits different in-gap states at structurally different kinds of step edges [30]. $UTe_2$ is also highly sensitive to the external control parameters such as pressure and magnetic field [19,31-34]. Moreover, re-entrant superconductivity is observed at high fields [35]. Remarkably, its superconducting ground state displays a large Sommerfeld coefficient ($\gamma$) [14,15,36], but without any corresponding thermal conductivity. However, this collection of remarkable properties have, to date, only been observed in the $UTe_2$ crystals with a nominally non-stoichiometric composition [36]. In this study, we employ the non-superconducting compound $UTe_2$, with nominally stoichiometric composition, as a target material to investigate the real composition of the obtained sample and the effect of pressure on its transport properties.

Our $UTe_2$ single crystals were grown using a vapor transport technique with a nominal ratio of 1U:2Te, as reported elsewhere [36]. Single crystal diffraction measurements indicate that our UTe$_2$ sample crystalizes in an orthorhombic unit cell with the lattice parameters $a$ =4.1615(0) Å, $b$=6.1203(0) Å, $c$=13.94300 (1) Å in the *Immm* (71) space group and all ions are fully occupied in the lattice (Fig. 1). To confirm the stoichiometry of our sample and clarify the argument whether the stoichiometry of the UTe$_2$ sample is associated with its superconductivity [14, 28, 37-40], we performed the chemical analysis by the technique of inductively coupled plasma atomic emission spectroscopy (ICP-AES), a direct method to detect the composition of material. We found that its actual composition is UTe$_{1.96}$, very close to the nominally stoichiometric UTe$_2$. Our chemical analysis results obtained from ICP-AES method is in good agreement with the result reported in 1996 by the same method [41]. Our consistent results from the analysis of the ICP-AES and the single crystal XRD measurements manifest that the sample investigated in this study is almost fully stoichiometric UTe$_2$.

## II. EXPERIMNTAL METHODS

### A. Setup for "quasi-uniaxial pressure experiment"

The uniaxial pressure shrinks the lattice along the $c$ axis but elongate the lattice along the $a$ or $b$ axes, while the hydrostatic pressure shrinks the lattice in all the three crystallographic directions equally. However, an ideal uniaxial pressure condition cannot be realized currently in the experimental measurement of the high-pressure resistance in a diamond anvil cell (DAC). There are two main reasons: one is that the high-pressure measurement in a DAC requires a metallic gasket to separate the two anvils, which can effectively avoid the damage of the anvils at high pressure due to that the gasket can serve as a buffer for the pressure applied; the other is that high-pressure resistance measurement in a DAC needs an insulating layer to prevent the conductance of the sample from the metallic gasket. In this study, we employed mixture of c-BN and epoxy as an insulating layer and fill them into the metallic gasket hole to protect the sample from short circuit with the metallic gasket. Then, the single crystal sample was put on the top of the insulating layer (Fig.2). In such a case, the compacted insulating powder can more or less restrain the lateral expansion along the $a$ or $b$ axes. Although our pressure environment cannot be the ideal uniaxial pressure that shrinks the lattice along the $c$ axis and elongate the lattice along the $a$ or $b$ axes, we can say that this is the available environment closest to the real uniaxial pressure in the high-pressure resistance measurements. Therefore, in order to appropriately or rigorously describe the pressure condition adopted in this study, we define the pressure applied on our sample as "quasi-uniaxial" pressure (see Fig. 2). To know the sample pressure, we placed a piece of ruby flake on the top of the sample. Since the ruby flake is in the same pressure environment as that of the sample, it can be used to characterize the pressure of the sample. Pressure was determined by the ruby fluorescence method [42].

**B. High-pressure resistance measurement along the *c* axis**

To obtain the effective information of the $c$-axis resistance from the pressurized samples, we designed the following experimental setup (Fig. 3). The closer the distance between the two electrodes on the *ab*-plane, the more favorable to minimize the effect of $R_{ab}(T)$. As a result, the distance between the electrodes is reduced to ~5μm.

## C. High-pressure susceptibility measurement

High pressure magnetic susceptibility measurements were carried out in a diamond anvil cell fabricated from Cu-Be alloy. The sample is surrounded by a secondary coil (pickup coil) and a field-generating primary coil which is wound on the top of the secondary coil. The alternating flux through the pickup coil produces an *ac* voltage which is the measured signal. When the sample is cooled below $T_c$, the field is expelled from the sample due to the superconducting shielding effect, forcing some of the flux lines out of the pickup coil and leading to a reduction in the induced voltage in the pickup coil [43-45]. For the primary coil, the alternating magnetic field was stimulated at 13.83 KHz, and the signal was collected at the same frequency. The superconducting transition can be captured by a step-like diamagnetic signal in this *ac* susceptibility measurement. The modulated magnetic field is provided by the secondary coil that is stimulated at 13.3 Hz, and its amplitude is around 9.9 mT. Here, the modulated susceptibility can be expressed by $\Delta\chi' = \chi'$ (9.9mT) - $\chi'$(0mT). Consequently, the result of $\chi'(B)$ produces a peak-like function in the secondary locked-in amplifier, instead of a step-like function which is collected by a single phase-locked amplifier. Such a measurement with the magnetic field modulation has been used by Geophysics lab in Carnegie Institute for the studies of pressure effect on superconductivity and pressure-induced superconductivity in materials [45-47].

## D. High-pressure specific heat measurement

For the high-pressure specific heat measurements, diamond anvils with 700 μm flat and non-magnetic rhenium gaskets were employed. A thin layered mixture of c-BN powder and epoxy was used as the insulating layer. The constantan was glued to one side of the sample with the dimension of about $0.4\times0.4\times0.05$ mm$^3$ as the heater; a chromel-AuFe (0.07%) thermocouple was fixed on the opposite side. With this technique, a small temperature oscillation (*ΔT*) generated by a heater is converted to an *ac* voltage signal. The *ac* calorimetry method adapted to high pressures was described in Refs. [48,49]. Pressure is determined by the ruby fluorescence method [42].

## III. RESULTS and DISCUSSIONS

### A. Resistance versus temperature along the *a*, *b* and *c* axes separately under quasi-uniaxial pressure and hydrostatic pressure

We applied the quasi-uniaxial pressure to the non-superconducting UTe$_2$ single crystals and performed comprehensive measurements of the temperature dependence of resistance, *R(T)* along the *a*, *b* and *c* axes as function of pressure. As shown in Figs. 4a, 4g and 4m, the transport properties at ambient-pressure in three crystallographic directions are highly anisotropic. From the fit of *R~log(T)*, we find that the resistance of our sample is linearly proportional to *log(T)* (Figs. 4a, 4g, 4m). This linear behavior indicates that the maxima in the resistivity is associated with an intrinsic energy scale (*T\**) of Kondo physics [50]. The coherence temperatures (*T\**) related to Kondo coherence (KC) are *T\**=88 K (Fig. 4g), 137 K (Fig. 4a) and 217 K (Fig. 4m) for currents in the *b*, *a* and *c*-directions respectively. Upon increasing pressure to 0.9 GPa, the measured resistance changes dramatically in all three directions, but a clear coherence temperature $T_b*(P)$ can only be resolved for current along the *b* axis. $T_b*(P)$ shifts to lower temperature upon compression (Fig. 3h), suggesting that the KC state is suppressed by pressure. An anisotropic response of the coherent temperature ($T^*$) of the UTe$_2$ single crystal to currents in the *a*, *b* and *c* directions is reminiscent of what is seen in the orthorhombic ferromagnetic superconductors UGe$_2$ [51], URhGe [52] and UCoGe [53]. It is likely that the layered material with an orthorhombic structure always holds the feature of the magnetic anisotropy. Such an anisotropy may result in the different $T^*$ values exhibited in the *R-T* curves along the three crystallographic axes, which calls for further investigations in the future.

Around 1.5 GPa, an abrupt drop in resistance develops in $R_a$ and $R_b$ upon cooling (Figs. 4c and 4i), while $R_c$ continues to grow on cooling (Fig. 4o). At low temperatures, $R_a(T)$ and $R_b(T)$ exhibit jumps in resistivity at ~3.5 K and ~2 K respectively (see insets of Fig.4c, 4i), features of which become more pronounced upon compression. At 2.8 GPa, a zero-resistance state appears in $R_a(T)$ and $R_b(T)$ (Figs. 4d and 4j), however, $R_c(T)$ still exhibits semiconducting behavior along the *c* axis (Fig. 4p). Finally, when the pressure reaches ~5 GPa, zero-resistance develops in all three directions (Figs. 4e, 4k and 4q). We repeated the measurements on new samples and found the results are reproducible (see Appendix).

We performed hydrostatic pressure measurements for the UTe$_2$ single crystal in the similar pressure range. As shown in Fig. 5, no superconducting transition was found under pressure up to 5.5 GPa for the first run of the experiments. To repeat the experimental results, we loaded the second sample into a diamond anvil cell and applied the hydrostatic pressure up to 6 GPa, no superconductivity is observed down to 0.3 K (see Appendix). These results indicate that the stochiometric UTe$_2$ is not superconducting under hydrostatic pressure condition below 6 GPa.

**B. Characterization of superconducting properties**

The two resistance jumps are closely correlated, leading us to tentatively attribute them to a superconducting ($T_c^{\text{onset}}$) and magnetic ($T_m$) transition. To confirm this hypothesis, we carried out three further sets of measurement. First, we applied magnetic fields under pressure at 2.4 GPa and 5.1 GPa, and found that the high-temperature resistance jump shifts to lower temperature (Figs. 6a-6d), while the second jump (knees) which lies between $T_c^{\text{onset}}$ and $T_c^{R=0}$ are not detectable above a field of 0.2 T, consistent with the identification that the jump in resistance is associated with a superconducting transition. Secondly, we simultaneously measured both *ac* susceptibility and resistance for the sample in the same high-pressure cell at 4.9 GPa (Fig. 6e). A sharp cusp in $\Delta\chi'(T)$ (green solids), indicating the development of diamagnetism, and zero resistance (red solids) were observed, in which the onset of diamagnetism in $\Delta\chi'(T)$ at 2.3 K coincides with the zero-resistance value [54]. Thirdly, we performed high-pressure specific heat (*C*) measurements on the sample at 6 GPa (Fig. 6f). Two peaks were observed in the temperature dependence of *C/T*. To clarify the origin of the two peaks, we applied a magnetic field perpendicular to the *ab*-plane and found that the high-temperature peak seen at 3.8 K moves to higher temperature and becomes quite broader at 1T. This behavior is similar to the feature of a FM phase,

*i.e.* its transition temperature in *C(T)* is broadened and softened with increasing magnetic fields [55,56]. Since the observed magnetic-like state in our sample has not been found before, further investigation to identify whether it is associated with an antiferromagnetic (AFM) or a FM state, or other exotic magnetic state is greatly needed. In this study, we define it as a magnetic (M) state. By contrast, the lower-temperature peak migrates downwards with increasing field, further confirming that it is associated with a zero-resistance superconducting transition ($T_c^{R=0}$). Moreover, the temperature of the low temperature peak detected by the specific heat is almost the same as that detected by resistance measurements, further supporting that it is resulted from a superconducting transition.

**C. High pressure X-ray diffraction**

To clarify whether the observed superconducting transition of the UTe$_2$ sample subjected to the quasi-uniaxial pressure is related to a pressure-induced crystal structure phase transition, we conducted the high-pressure synchrotron X-ray diffraction measurements on the powder sample at 20 K at beamline 4W2 of the Beijing Synchrotron Radiation Facility. No new peak was observed and all peaks can be indexed well by the orthorhombic structure with the space group *Immm* up to 6.1 GPa (Fig. 7). These results indicate that no crystal structure phase transition occurs, providing the useful structure information for the pressure condition under which the resistance, susceptibility and heat capacity are measured. Since the quasi-uniaxial pressure environment is the combination of the uniaxial pressure with hydrostatic pressure, isolating the combined pressure is quite helpful to reveal the true pressure effect on the lattice parameters and deserves investigations in the future, although it is difficult in practical experiments.

**D. Pressure-temperature phase diagram for the compressed UTe$_2$ single crystal**

We summarize our experimental results in the pressure-temperature phase diagram in Fig.8. It is seen that the onset temperature ($T^*$) of the Kondo coherence is suppressed by pressure and is not detectable at ~ 1.5 GPa, the pressure where the superconductivity emerges on the *ab*-plane at 3.6 K, followed by the exotic magnetic (M) phase transition (Fig. 4 and Fig.6). The superconducting state, featured by the upturn of resistance in *c* direction, coexists with the M state below 2.8 GPa (see Appendix), and the onset transition temperature of these two states ($T_c^{onset}$ and $T_M$) varies slowly with increasing pressure. At the critical pressure $P_{c1}$ (2.8 GPa), zero resistance of the superconducting state of the *ab*-plane is observed around 0.9 K (Figs. 4d and 4j) and rising to ~2 K at 4.8 GPa ($P_{c2}$), where the superconducting state develops in the *c* direction. The zero resistance $T_c$ of the superconducting state measured in all crystallographic directions for different experimental runs varies between 2 K and 1.8 K (Figs. 4f-4r and Appendix).

Unexpectedly, we find that the magnetism develops just below the onset $T_c$ of the superconducting state, suggesting that the superconducting transition gives rise to a M state. This contrasts with the other superconductors, in which the superconductivity emerges from a pre-existing but suppressed M state [57-59]. We noted the work, reported by Thomas et al [60], that the pressure-induced magnetic ordered state is different from ours. For our study, the magnetic ordered state in UTe$_2$ emerges from the superconducting state appeared just below the onset temperature of the superconducting transition (Fig.8), while a magnetic state presents quite differently under the hydrostatic pressure condition where the superconducting state is suppressed [60]. We presume that the difference may stem from the two aspects: the composition of the sample and the pressure environment applied - the composition of their sample is UTe$_2$ with unneglectable amount of Te deficiency and the pressure environment is hydrostatic, while the composition of our sample is almost stoichiometric UTe$_2$ (UTe$_{1.96}$) and the pressure environment is quasi-uniaxial. In the case of the quasi-uniaxial pressure, the sample tends to shrink more along the *c* axis but very less along the *a* and *b* axes, while in the case of the hydrostatic pressure, the sample is supposed to be compressed equally in all the three crystallographic directions. Such distinct pressure effects between the two experiments may impact differently on the easy axis of its magnetic ordered structure, which may lead to a diverse magnetic state.

Our results raise several questions: (1) Is the stoichiometric UTe$_2$ the parent compound of superconducting UTe$_{2-\delta}$? (2) Is the role of Te deficiency similar to or different from the other unconventional superconductors with anion deficiencies [7,61-63]? (3) Does the pressure-induced superconductivity from the stoichiometric UTe$_2$ has a nontrivially topological nature? All these questions are the attractive issues which deserve further investigations.

## IV. CONCLUSION

We report the observation of quasi-uniaxial-pressure-induced superconductivity in stoichiometric compound UTe$_2$ that is not superconducting at ambient pressure. The measurements of resistivity, magnetoresistance, specific heat and susceptibility reveal that the quasi-uniaxial pressure can suppress the Kondo coherent state and then drive a superconducting transition initially emerged on the *ab*-plane at 1.5 GPa. At 4.8 GPa, the superconductivity is developed in three crystallographic directions. The superconducting state coexists with an exotic magnetic ordered state that presents just below the onset temperature of the superconducting transition. These results highlight the vital role of Te deficiency in developing the superconductivity at ambient pressures, and also shed new insight on understandings on the underlying superconducting mechanisms in the correlated electron systems of the UTe$_2$, and even in the other unconventional superconducting compounds.

## V. APENDIX

### 1. Reproducible resistance measurements on our UTe$_2$ single crystal under quasi-uniaxial pressure

We performed the same high-pressure measurements on the UTe$_2$ single crystals that were cut from different batches. As shown in Fig. A1, $R_a$ and $R_b$ as a function of temperature for the sample 2 exhibit a similar behavior to that of the sample 1 (Fig.4). Since the thickness of the sample 2 is thicker than that of the sample 1, we observed a flat residual resistance in $R_b(T)$ starting at ~ 1.6 K on cooling at 3.7 GPa. It is known that the observed resistance ($R$) is composed of three parts, sample resistance ($R_s$), contact resistance between the sample and electrodes ($R_c$) and deformation resistance ($R_d$) which is associated with cracks, *i.e.* $R=R_s+R_c+R_d$. The last two terms of the observed resistance are related to residual resistance ($R_r$, $R_r= R_c+R_d$). If the sample is brittle and thicker, it is easier to be cracked by pressure. In this case, $R_d>0$ and it gives rise to $R_r$. T-independent behavior with $R_r \sim$ 0.015 Ohm at 3.7 GPa starting at ~1.6 K (Figs. A1c and A1g), and 0.0085 Ohm at 4.8 GPa starting at ~2 K (Figs. A1d and A1h) indicates superconducting transition because no superconductivity was observed from the same electrodes at temperature down to 20 mK in the megabar pressure range [64].

In order to further confirm the pressure-induced superconducting and M phase transition for the nominally stoichiometric $UTe_2$, we conducted resistance measurements on the sample that were cut from different batches for the third run and obtained the very similar results (Fig. A2).

To display the linear behavior in *log(T)* clearly, we fit the data equally by a straight line for Figs.2a, 2g and 2m (see Fig. A3).

## 2  Reproducible resistance measurements on our $UTe_2$ single crystal under hydrostatic pressure

To repeat the experimental results, we loaded the second sample into a diamond anvil cell and applied the hydrostatic pressure up to 6 GPa, no superconductivity is observed down to 0.3 K (Fig. A4). These results indicate that the stochiometric $UTe_2$ is not superconducting under hydrostatic pressure condition below 6 GPa.

**Acknowledgements**

The work in China was supported by the National Key Research and Development Program of China (Grant No. 2021YFA1401800, 2017YFA0302900 and 2017YFA0303100), the NSF of China (Grants No. U2032214, 11888101, 12122414, 12004419, 12104487 and 11874330), the Strategic Priority Research Program (B) of the Chinese Academy of Sciences (Grant No. XDB25000000), the Fund of Science and Technology on Surface Physics and Chemistry Laboratory (201901SY00900102). We thank the support from the Users with Excellence Program of Hefei Science Center CAS (2020HSC-UE015). Part of the work is supported by the Synergic Extreme Condition User System. J. G. and S.C. are grateful for supports from the Youth Innovation Promotion Association of the CAS (2019008) and the China Postdoctoral Science Foundation (E0BK111). Piers Coleman is supported by the US National Science Foundation grant DMR-1830707.

Correspondence and requests for materials should be addressed to L.S.(llsun@iphy.ac.cn).

These authors with star (*) contributed equally to this work.

**References**

[1] P. Coleman and A. J. Schofield, Nature **433**, 226 (2005).

[2] B. Keimer, S. A. Kivelson, M. R. Norman, S. Uchida and J. Zaanen, Nature **518**, 179 (2015).

[3] L. Sun, X. J. Chen, J. Guo, P. Gao, Q. Z. Huang, H. Wang, M. Fang, X. Chen, G. Chen, Q. Wu, C. Zhang, D. Gu, X. Dong, L. Wang, K. Yang, A. Li, X. Dai, H. K.

Mao and Z. Zhao, Nature **483**, 67 (2012).

[4] N. D. Mathur, F. M. Grosche, S. R. Julian, I. R. Walker, D. M. Freye, R. K. W. Haselwimmer and G. G. Lonzarich, Nature **394**, 39 (1998).

[5] S. Uji, H. Shinagawa, T. Terashima, T. Yakabe, Y. Terai, M. Tokumoto, A. Kobayashi, H. Tanaka and H. Kobayashi, Nature **410**, 908 (2001).

[6] H. Mukuda, M. Abe, Y. Araki, Y. Kitaoka, K. Tokiwa, T. Watanabe, A. Iyo, H. Kito and Y. Tanaka, Phys. Rev. Lett. **96**, 087001 (2006).

[7] C. Proust and L. Taillefer, Annu. Rev. Condens. Matter Phys. **10**, 409 (2019).

[8] J. Zaanen, SciPost Physics **6**, 061 (2019).

[9] R. L. Greene, P. R. Mandal, N. R. Poniatowski and T. Sarkar, Ann. Rev. Condens. Matter Phys. **11**, 213 (2020).

[10] E. Dagotto, Rev. Mod. Phys. **85**, 849-867 (2013).

[11] B. Shen, Y. Zhang, Y. Komijani, M. Nicklas, R. Borth, A. Wang, Y. Chen, Z. Nie, R. Li, X. Lu, H. Lee, M. Smidman, F. Steglich, P. Coleman and H. Yuan, Nature **579**, 51 (2020).

[12] T. Park, F. Ronning, H. Q. Yuan, M. B. Salamon, R. Movshovich, J. L. Sarrao and J. D. Thompson, Nature **440**, 65 (2006).

[13] D. Aoki and J. Flouquet, J. Phys. Soc. Jpn. **81**, 011003 (2011).

[14] S. Ran, C. Eckberg, Q.-P. Ding, Y. Furukawa, T. Metz, S. R. Saha, I. L. Liu, M. Zic, H. Kim, J. Paglione and N. P. Butch, Science **365**, 684 (2019).

[15] D. Aoki, A. Nakamura, F. Honda, D. Li, Y. Homma, Y. Shimizu, Y. J. Sato, G. Knebel, J.-P. Brison, A. Pourret, D. Braithwaite, G. Lapertot, Q. Niu, M. Vališka, H. Harima and J. Flouquet, J. Phys. Soc. Jpn. **88**, 043702 (2019).

[16] Y. Xu, Y. Sheng and Y. F. Yang, Phys. Rev. Lett. **123**, 217002 (2019).

[17] S.-i. Fujimori, I. Kawasaki, Y. Takeda, H. Yamagami, A. Nakamura, Y. Homma and D. Aoki, J. Phys. Soc. Jpn. **88**, 103701 (2019).

[18] L. Miao, S. Liu, Y. Xu, E. C. Kotta, C. J. Kang, S. Ran, J. Paglione, G. Kotliar, N. P. Butch, J. D. Denlinger and L. A. Wray, Phys. Rev. Lett. **124**, 076401 (2020).

[19] S. M. Thomas, F. B. Santos, M. H. Christensen, T. Asaba, F. Ronning, J. D. Thompson, E. D. Bauer, R. M. Fernandes, G. Fabbris and P. F. S. Rosa, Sci. Adv.

**6**, eabc8709 (2020).

[20] S. Sundar, S. Gheidi, K. Akintola, A. M. Côté, S. R. Dunsiger, S. Ran, N. P. Butch, S. R. Saha, J. Paglione and J. E. Sonier, Phys. Rev. B **100**, 140502(R) (2019).

[21] Y. Tokunaga, H. Sakai, S. Kambe, T. Hattori, N. Higa, G. Nakamine, S. Kitagawa, K. Ishida, A. Nakamura, Y. Shimizu, Y. Homma, D. Li, F. Honda and D. Aoki, J. Phys. Soc. Jpn. **88**, 073701 (2019).

[22] C. Duan, K. Sasmal, M. B. Maple, A. Podlesnyak, J.-X. Zhu, Q. Si and P. Dai, Phys. Rev. Lett. **125**, 237003 (2020).

[23] C. Duan, R. E. Baumbach, A. Podlesnyak, Y. Deng, C. Moir, A. J. Breindel, M. B. Maple, E. M. Nica, Q. Si and P. Dai, Nature **600**, 636 (2021).

[24] W. Knafo, G. Knebel, P. Steffens, K. Kaneko, A. Rosuel, J. P. Brison, J. Flouquet, D. Aoki, G. Lapertot and S. Raymond, Phys. Rev. B **104**, L100409 (2021).

[25] M. Sato and Y. Ando, Rep. Prog. Phys. **80**, 076501 (2017).

[26] G. Nakamine, S. Kitagawa, K. Ishida, Y. Tokunaga, H. Sakai, S. Kambe, A. Nakamura, Y. Shimizu, Y. Homma, D. Li, F. Honda and D. Aoki, J. Phys. Soc. Jpn. **88**, 113703 (2019).

[27] I. M. Hayes, D. S. Wei, T. Metz, J. Zhang, Y. S. Eo, S. Ran, S. R. Saha, J. Collini, N. P. Butch, D. F. Agterberg, A. Kapitulnik and P. Johnpierre, arXiv:2002.02539 (2020).

[28] D. Aoki, A. Nakamura, F. Honda, D. Li, Y. Homma, Y. Shimizu, Y. J. Sato, G. Knebel, J.-P. Brison, A. Pourret, D. Braithwaite, G. Lapertot, Q. Niu, M. Vališka, H. Harima and J. Flouquet, JPS Conf. Proc. **30**, 011065 (2020).

[29] J. L. Lado and P. Liljeroth, A layered unconventional superconductor, Nat. Phys. **17**, 1287 (2021).

[30] L. Jiao, S. Howard, S. Ran, Z. Wang, J. O. Rodriguez, M. Sigrist, Z. Wang, N. P. Butch and V. Madhavan, Nature **579**, 523 (2020).

[31] S. Ran, H. Kim, I. L. Liu, S. R. Saha, I. Hayes, T. Metz, Y. S. Eo, J. Paglione and N. P. Butch, Phys. Rev. B **101**, 140503(R) (2020).

[32] D. Braithwaite, M. Vališka, G. Knebel, G. Lapertot, J. P. Brison, A. Pourret, M. E.

Zhitomirsky, J. Flouquet, F. Honda and D. Aoki, Commun. Phys. **2**, 147 (2019).

[33] D. Aoki, F. Honda, G. Knebel, D. Braithwaite, A. Nakamura, D. Li, Y. Homma, Y. Shimizu, Y. J. Sato, J.-P. Brison and J. Flouquet, J. Phys. Soc. Jpn. **89**, 053705 (2020).

[34] W.-C. Lin, D. J. Campbell, S. Ran, I. L. Liu, H. Kim, A. H. Nevidomskyy, D. Graf, N. P. Butch and J. Paglione, Npj Quantum Mater. **5**, 68 (2020).

[35] S. Ran, I. L. Liu, Y. S. Eo, D. J. Campbell, P. M. Neves, W. T. Fuhrman, S. R. Saha, C. Eckberg, H. Kim, D. Graf, F. Balakirev, J. Singleton, J. Paglione and N. P. Butch, Nat. Phys. **15**, 1250 (2019).

[36] S. Ikeda, H. Sakai, D. Aoki, Y. Homma, E. Yamamoto, A. Nakamura, Y. Shiokawa, Y. Haga and Y. Ōnuki, J. Phys. Soc. Jpn. **75**, 116 (2006).

[37] L. P. Cairns, C. R. Stevens, C. D. O'Neill and A. Huxley, J. Phys. Condens. Matter. **32**, 415602 (2020).

[38] P. F. S. Rosa, A. Weiland, S. S. Fender, B. L. Scott, F. Ronning, J. D. Thompson, E. D. Bauer and S. M. Thomas, arXiv:2110.06200 (2021).

[39] S. M. Thomas, C. Stevens, F. B. Santos, S. S. Fender, E. D. Bauer, F. Ronning, J. D. Thompson, A. Huxley and P. F. S. Rosa, Phys. Rev. B **104**, 224501 (2021).

[40] Y. Haga, P. Opletal, Y. Tokiwa, E. Yamamoto, Y. Tokunaga, S. Kambe and H. Sakai, J. Phys. Condens. Matter. **34**, 175601 (2022).

[41] K. Stowe, J. of solid Chem. **127**, 202 (1996).

[42] H. K. Mao, J. Xu, P. M. Bell, J. Geophys. Res. **91**, 4673 (1986).

[43] J. J. Hamlin, V. G. Tissen, J. S. Schilling, Phys. Rev. B **73**, 094522 (2006).

[44] L. L. Sun, X.-J. Chen, J. Guo, P. W. Gao, Q.-Z. Huang, H. D. Wang, M. H. Fang, X. L. Chen, G. F. Chen, Q. Wu, C. Zhang, D. C. Gu, X. L. Dong, L. Wang, K. Yang, A. G. Li, X. Dai, H.-k. Mao, Z. X. Zhao, Nature **483**, 67 (2012).

[45] Y. A. Timofeev, V. V. Struzhkin, R. J. Hemley, H.-k. Mao, E. A. Gregoryanz, Rev. Sci. Instrum **73**, 371 (2002).

[46] V. V. Struzhkin, R. J. Hemley, H.-k. Mao, Y. A. Timofeev, Nature **390**, 382 (1997).

[47] X.-J. Chen, V. V. Struzhkin, Y. Yu, A. F. Goncharov, C.-T. Lin, H.-k. Mao, R. J.

Hemley, Nature **466**, 950 (2010).

[48] A. Eichler, W. Gey, Rev. Sci. Instrum **50**, 1445 (1979).

[49] V. A. Sidorov, J. D. Thompson, and Z. Fisk, J. Phys.: Condensed Matter **22**, 406002 (2010).

[50] M. C. Aronson, J. D. Thompson, J. L. Smith, Z. Fisk and M. W. McElfresh, Phys. Rev. Lett. **63**, 2311 (1989).

[51] I. Sheikin, A. Huxley, D. Braithwaite, J. P. Brison, S. Watanabe, K. Miyake and J. Flouquet, Phys. Rev. B **64**, 220503 (2001).

[52] F. Hardy and A. D. Huxley, Phys. Rev. Lett. **94**, 247006 (2005).

[53] N. T. Huy, D. E. de Nijs, Y. K. Huang and A. de Visser, Phys. Rev. Lett. 100, 077002 (2008).

[54] J. Guo, Y. Zhou, C. Huang, S. Cai, Y. Sheng, G. Gu, C. Yang, G. Lin, K. Yang, A. Li, Q. Wu, T. Xiang and L. Sun, Nat. Phys. **16**, 295 (2019).

[55] S. Uhlenbruck, R. Teipen, R. Klingeler, B. Büchner, O. Friedt, M. Hücker, H. Kierspel, T. Niemöller, L. Pinsard, A. Revcolevschi and R. Gross, Phys. Rev. Lett. **82**, 185 (1999).

[56] P. Boulet, E. Colineau, F. Wastin, P. Javorský, J. C. Griveau, J. Rebizant, G. R. Stewart and E. D. Bauer, Phys. Rev. B **72**, 064438 (2005).

[57] S. S. Saxena, P. Agarwal, K. Ahilan, F. M. Grosche, R. K. W. Haselwimmer, M. J. Steiner, E. Pugh, I. R. Walker, S. R. Julian, P. Monthoux, G. G. Lonzarich, A. Huxley, I. Sheikin, D. Braithwaite and J. Flouquet, Nature **406**, 587-592 (2000).

[58] D. Aoki, A. Huxley, E. Ressouche, D. Braithwaite, J. Flouquet, J.-P. Brison, E. Lhotel and C. Paulsen, Nature **413**, 613 (2001).

[59] N. T. Huy, A. Gasparini, D. E. de Nijs, Y. Huang, J. C. P. Klaasse, T. Gortenmulder, A. de Visser, A. Hamann, T. Görlach and H. v. Löhneysen, Phys. Rev. Lett. **99**, 067006 (2007).

[60] S. M. Thomas, F. B. Santos, M. H. Christensen, T. Asaba, F. Ronning, J. D. Thompson, E. D. Bauer, R. M. Fernandes, G. Fabbris and P. F. S. Rosa, Science Advances **6**, eabc8709 (2020).

[61] R. Zhi-An, L. Wei, Y. Jie, Y. Wei, S. Xiao-Li, C. Zheng, C. Guang-Can, D.

Xiao-Li, S. Li-Ling, Z. Fang and Z. Zhong-Xian, Chin. Phys. Lett. **25**, 2215 (2008).

[62] X. C. Wang, Q. Q. Liu, Y. X. Lv, W. B. Gao, L. X. Yang, R. C. Yu, F. Y. Li and C. Q. Jin, Solid State Commun. **148**, 538 (2008).

[63] T. K. Chen, C. C. Chang, H. H. Chang, A. H. Fang, C. H. Wang, W. H. Chao, C. M. Tseng, Y. C. Lee, Y. R. Wu, M. H. Wen, H. Y. Tang, F. R. Chen, M. J. Wang, M. K. Wu and D. Van Dyck, PNAS **111**, 63 (2014).

[64] L. Sun, T. Matsuoka, Y. Tamari, K. Shimizu, J. Tian, Y. Tian, C. Zhang, C. Shen, W. Yi, H. Gao, J. Li, X. Dong and Z. Zhao, Phys. Rev. B **79**, 140505(R) (2009).

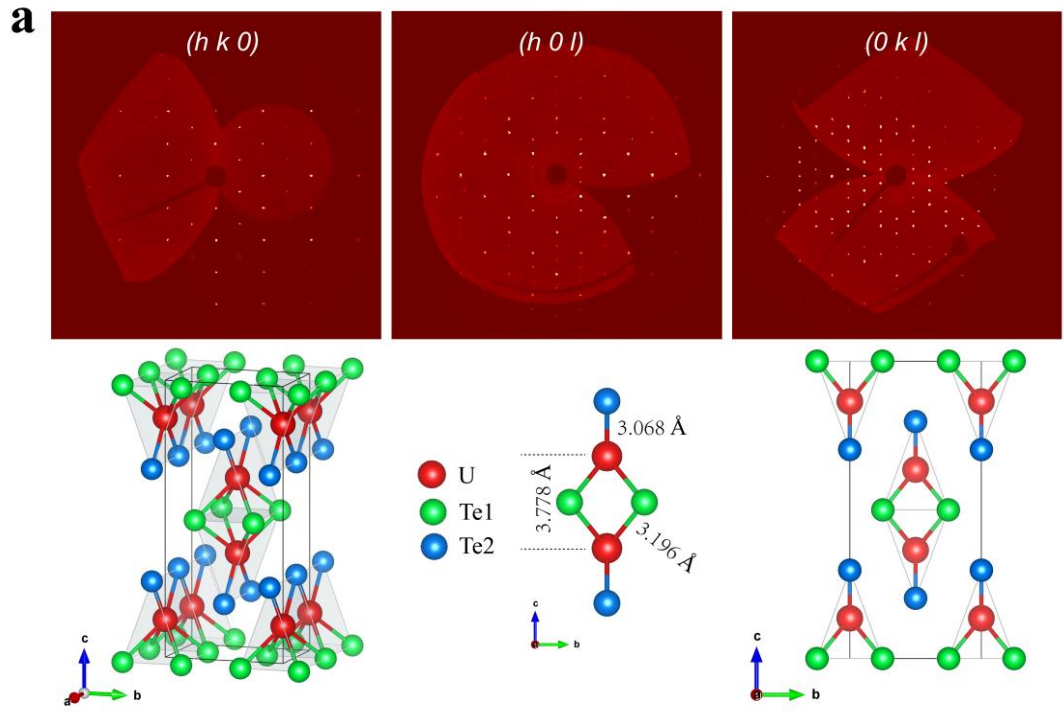

Figure 1 Crystal structure and composition information for the nominally stoichiometric UTe$_2$ employed in the present study. (a) The single crystal x-ray diffraction patterns for the (*hk0*), (*h0l*) and (*0kl*) zones taken at 300 K. (b) The lattice and structure parameters of our UTe$_2$ sample.

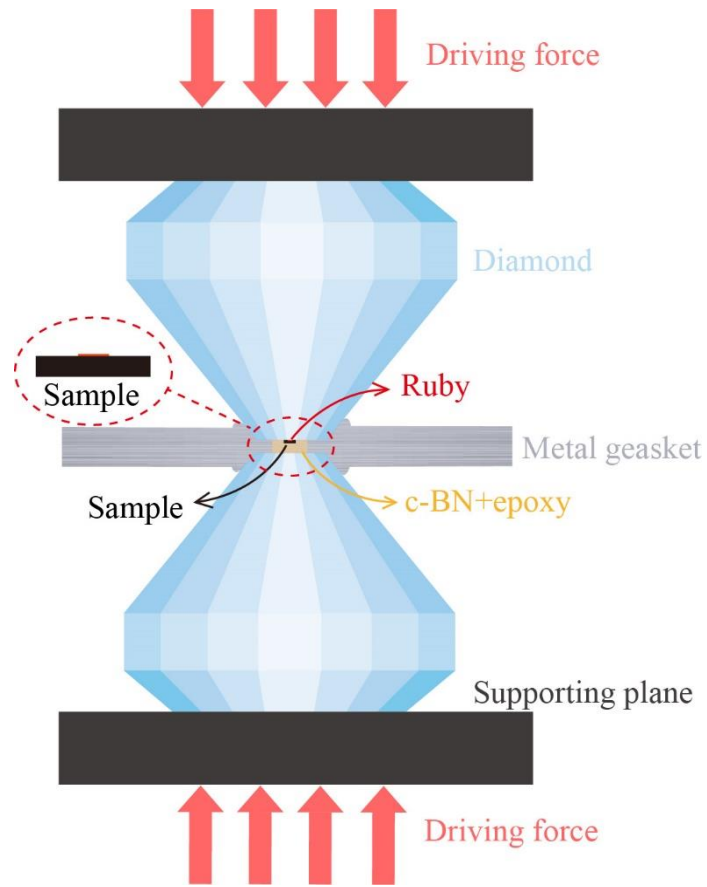

Figure 2 A sketch of our quasi-uniaxial pressure experiment setup, displaying the arrangement for the gasket (gray), c-BN insulating layer (yellow), sample (black) and ruby (red) in a diamond anvil cell. The circle on the left displays the enlarged view for the sample and ruby flake.

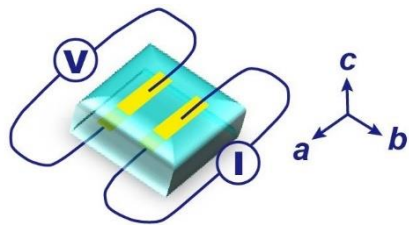

Figure 3 Experimental setup for the *c*-axis resistance measurements. The bule stands for the sample. The yellow and green bars represent the electrodes on the top and the bottom of the sample, respectively. The distance between the electrodes is about 5μm.

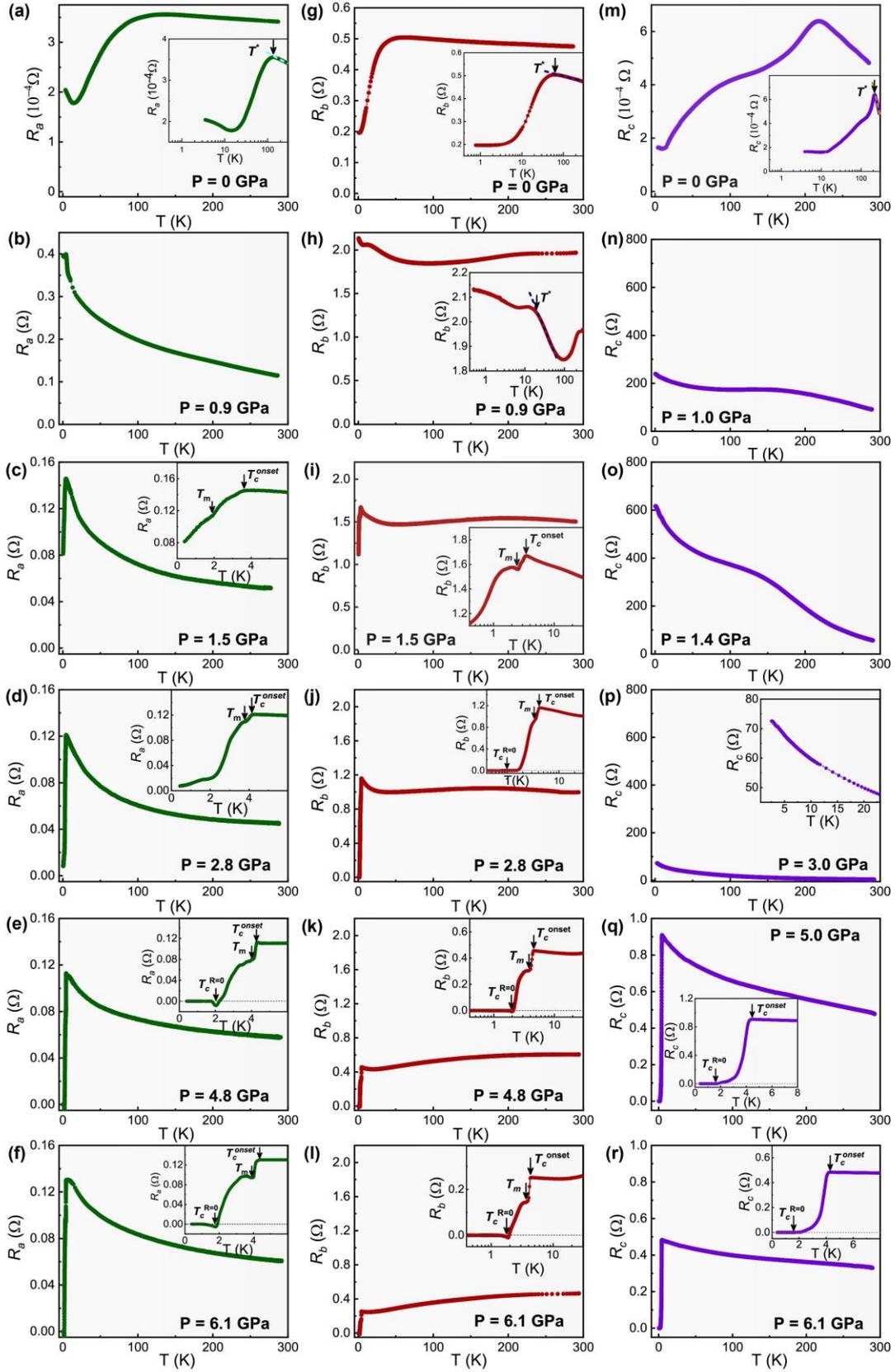

Figure 4 Characterization of transport properties of single crystal UTe$_2$. Temperature dependence of electrical resistance $R(T)$, measured along the *a* axis (a-f), *b* axis (g-l)

and *c* axis (m-r), between 1 bar and 6.1 GPa. The insets detail the low temperature behavior. $T^*$ and $T_m$ denote the onset temperature of the Kondo coherence and magnetic phase transitions. $T_c^{onset}$ and $T_c^{R=0}$ stand for the temperatures of onset and zero-resistance superconducting phase transitions, respectively.

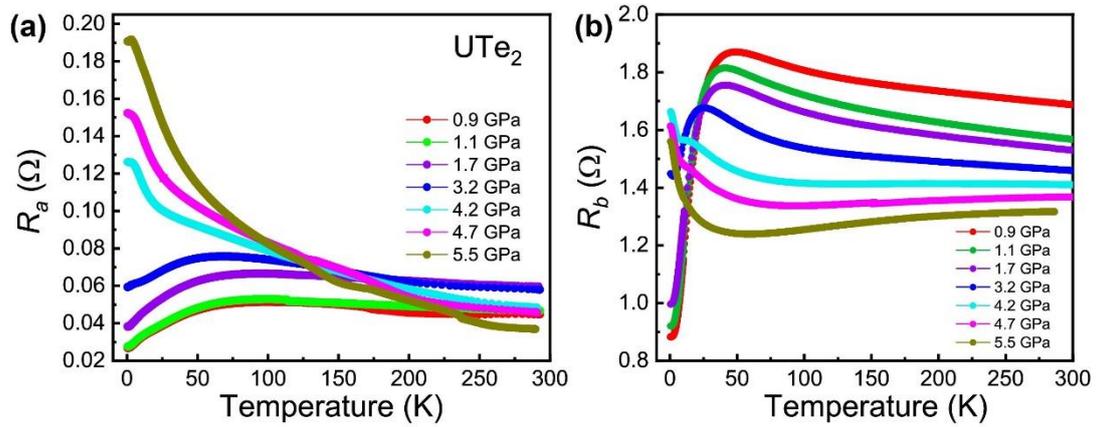

Figure 5 Temperature dependence of $R_a$ and $R_b$ of the UTe$_2$ sample measured under hydrostatic pressure conduction, showing no superconducting transition up to 5.5 GPa.

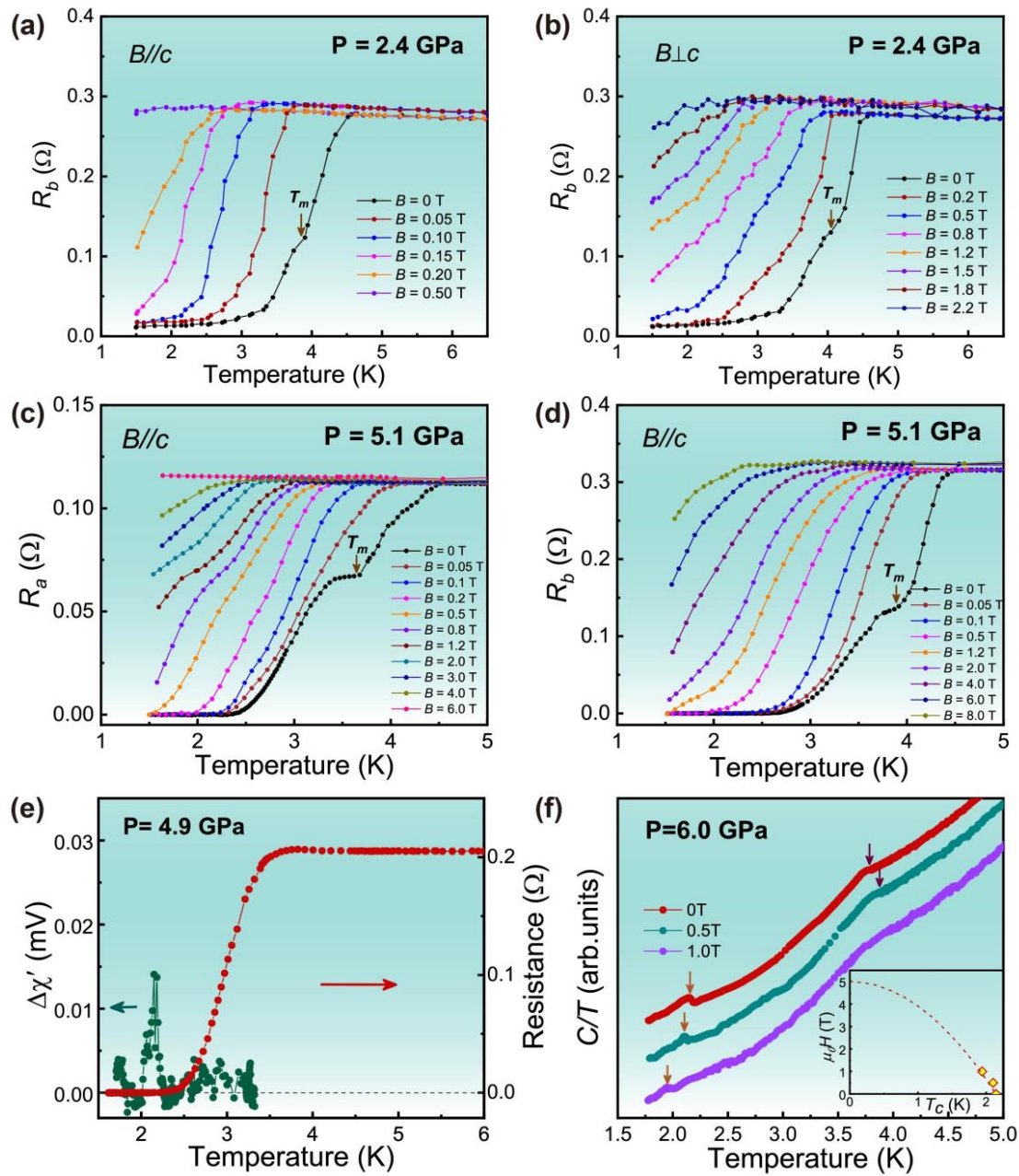

Figure 6 Characterization of superconducting properties in single crystal UTe$_2$. (a-b) The temperature dependence of resistance along $b$ axis ($R_b$) at 2.4 GPa under different magnetic fields, showing a dramatic dependence on the direction of the applied field. (c-d) The resistance along the $a$ and $b$ axes ($R_a$ and $R_b$) as a function of temperature under different magnetic fields at 5.1 GPa. (e) Planar resistance and $ac$ susceptibility ($\Delta\chi'$) as a function of temperature. Green solid points are the $ac$ susceptibility, while red points show the resistance. (f) Specific heat coefficient ($C/T$) with respect to temperature (*T)* measured at 6 GPa under different fields. The lower temperature peak is associated with a bulk superconducting transition, while the higher temperature peak is magnetic. The inset shows the plot of superconducting transition temperature $T_c$ versus critical field ($H_{c2}$). The dashed lines represent the Ginzburg-Landau fit to the data of $H_{c2}$.

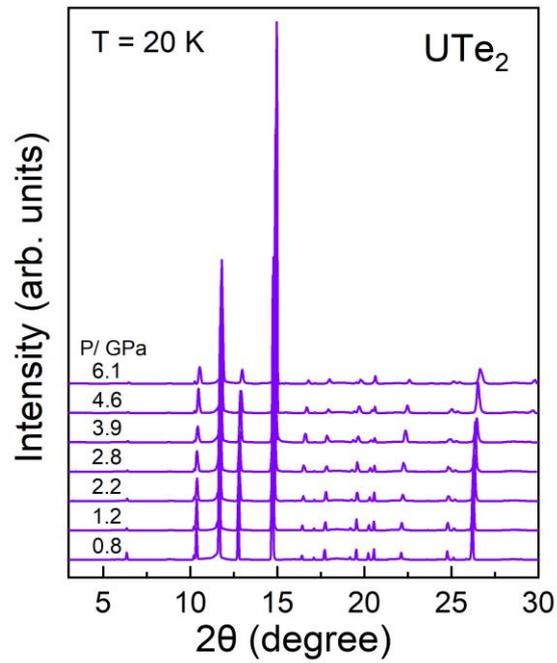

Figure 7 X-ray diffraction patterns of the UTe$_2$ sample collected at 20 K and different pressures.

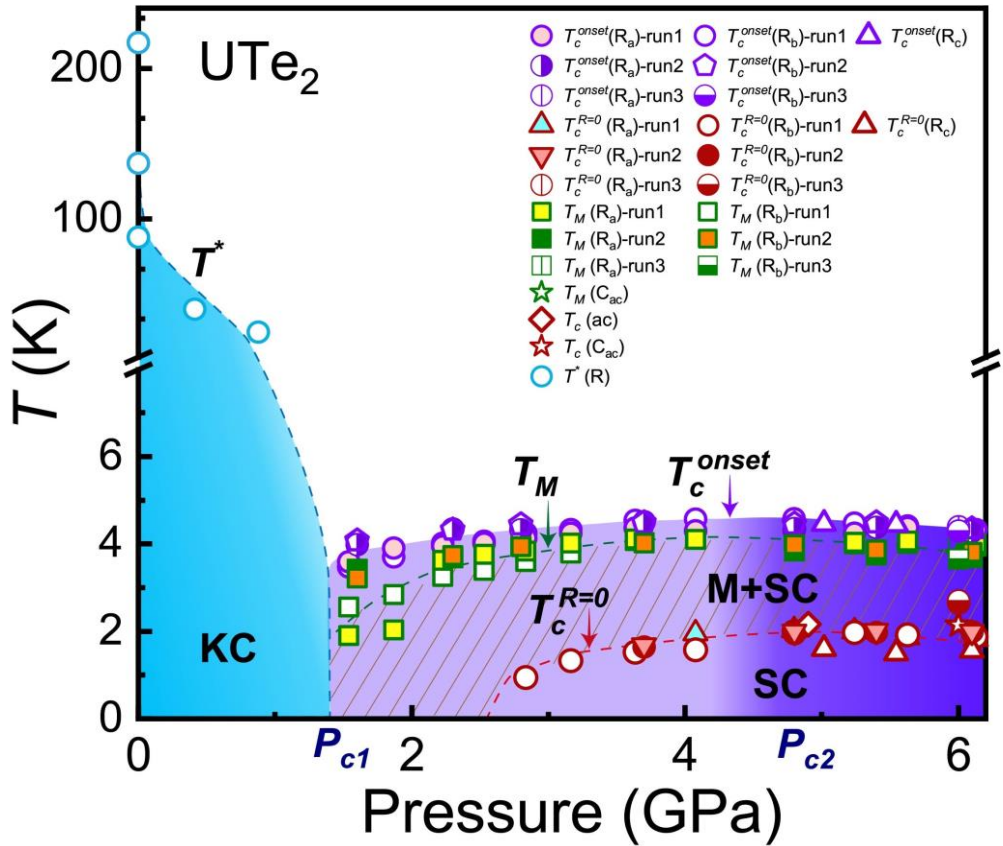

Figure 8 Pressure-temperature phase diagram for the compressed UTe$_2$ single crystal. The evolution of electronic state under uniaxial pressure, demonstrating the emergence of the superconducting and the exotic magnetic (M) state at $P_{c1}$ (1.5 GPa) from a Kondo coherence (KC) state, and a crossover from planar superconductivity to a bulk superconducting state at $P_{c2}$ (4.8 GPa). $T^*$ represents the onset temperature of the KC state, $T_c^{onset}(R_a)$, $T_c^{onset}(R_b)$ and $T_c^{onset}(R_c)$ represent the onset temperature of the superconducting transition detected by resistance measurements for currents in the $a$, $b$ and $c$ directions. $T_c^{R=0}(R_a)$, $T_c^{R=0}(R_b)$ and $T_c^{R=0}(R_c)$ stand for the zero-resistance superconducting transition temperature detected by resistance measurements for currents in the $a$, $b$ and $c$ directions. $T_M(R)$ and $T_M(C_{ac})$ stand for the M phase transition temperature obtained by resistance and heat capacity

measurements. $T_c(ac)$ and $T_c$ ($C_{ac}$) represent the superconducting transition temperature measured through *ac* susceptibility and specific heat measurements. The SC region filled with light violet refers to the superconductivity realized on the *ab*-plane only, while the region filled with dark violet refers to the superconductivity realized in all three crystallographic directions.

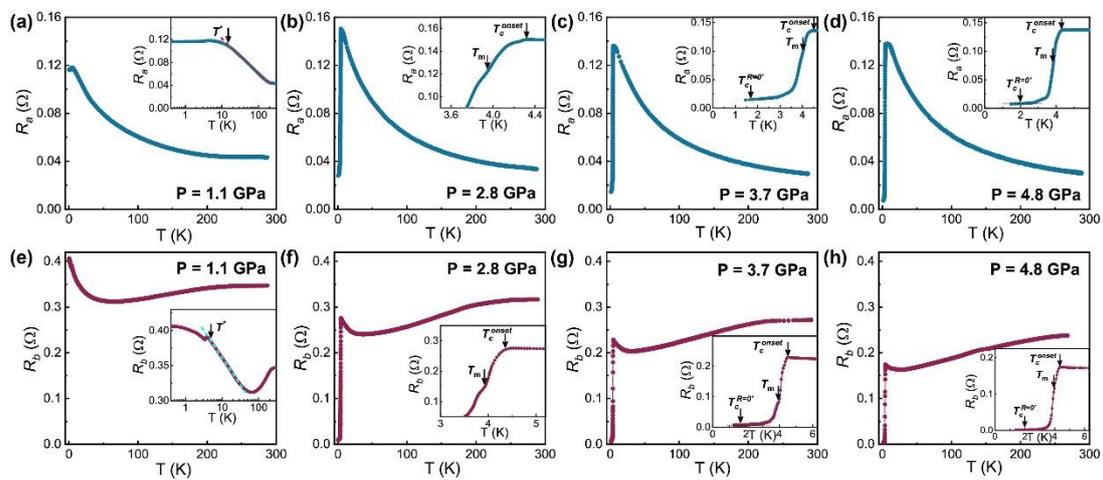

Fig. A1. Temperature dependence of $R_a$ and $R_b$ for the sample 2 at different pressures.

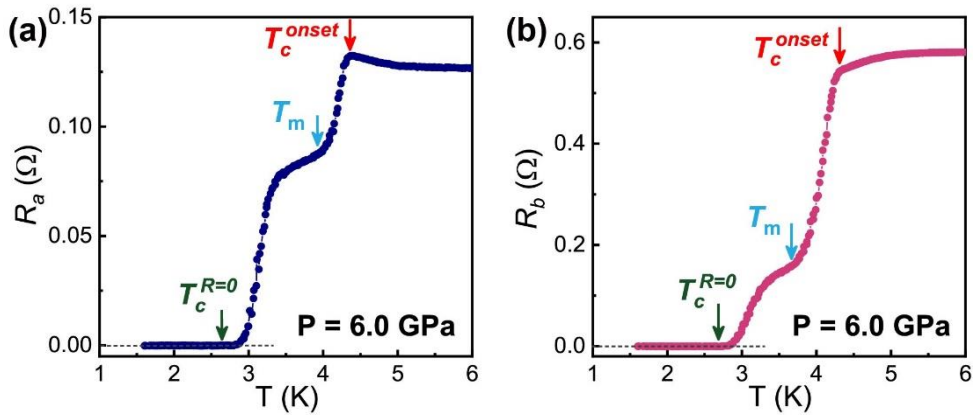

Fig. A2 Characterization of superconducting and M phase transition for the single crystal UTe$_2$ that is subjected to 6 GPa.

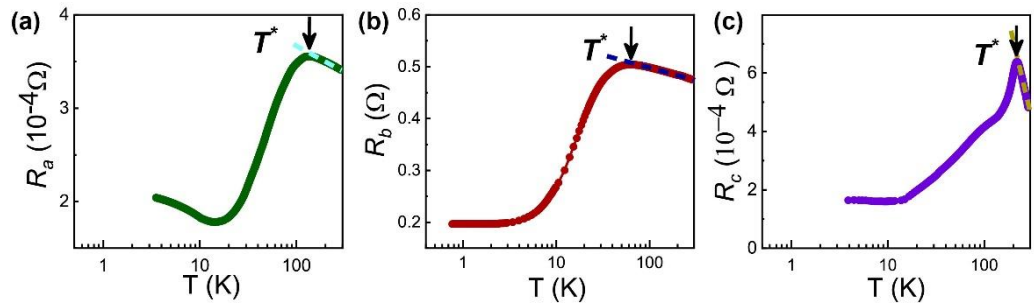

Fig. A3 Temperature dependence of resistance in three crystallographic directions, displaying the enlarged view of Fig. 4a, 4g and 4m.

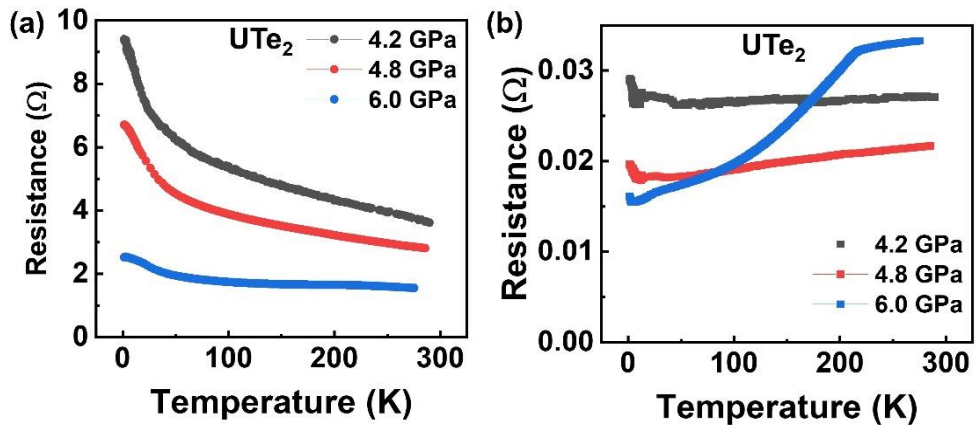

Fig. A4 $R_a$ and $R_b$ as a function of temperature for the UTe$_2$ sample measured under hydrostatic pressure condition up to 6 GPa.